\documentclass[11pt]{article}

\usepackage{graphicx}
\usepackage{amsmath, amsfonts, amssymb, mathrsfs, rotating, setspace, amsfonts}
\usepackage{natbib}
\usepackage[left=1in,top=1in,right=1in,bottom=1in]{geometry} 
\usepackage[justification=centering]{caption}

\title{Comparing Biomarkers as Trial Level General Surrogates}

\author{Erin E Gabriel\\
Biostatistics Research Branch, Division of Clinical Research, NIAID/NIH, Bethesda, MD USA\\
Michael J Daniels\\
Department of Statistics and Data Sciences, The University of Texas at Austin  USA\\
Department of Integrative Biology, The University of Texas at Austin  USA\\
M Elizabeth Halloran\\
Department of Biostatistics, School of Public Health, University of Washington, Seattle, WA, USA\\
Vaccine and Infectious Diseases Division, Fred Hutchinson Cancer Research Center, Seattle, WA, USA\\
Center for Inference and Dynamics of Infectious Diseases, Seattle, WA, USA}

\begin{document}
\maketitle

\begin{abstract}
An intermediate response measure that accurately predicts efficacy in a new setting can reduce trial cost and time to product licensure. In this paper, we define a \textit{trial level general surrogate} as a trial level intermediate response that accurately predicts trial level clinical responses. Methods for evaluating trial level general surrogates have been developed previously. Many methods in the literature use trial level intermediate responses for prediction. However, all existing methods focus on surrogate evaluation and prediction in new settings, rather than comparison of candidate trial level surrogates, and few formalize the use of cross validation to quantify the expected prediction error. Our proposed method uses Bayesian non-parametric  modeling and cross-validation to estimate the absolute prediction error for use in evaluating and comparing candidate trial level general surrogates. Simulations show that our method performs well across a variety of scenarios. We use our method to evaluate and to compare candidate trial level general surrogates in several multi-national trials of a pentavalent rotavirus vaccine. We identify two immune measures that have potential value as trial level general surrogates and use the measures to predict efficacy in a trial with no clinical outcomes measured. \\
Bayesian non-parametrics; Cross-Validation; Meta-analysis; Surrogate markers; Vaccines
\end{abstract}

\newpage

\linespread{2}
\section{Introduction}
Many different definitions of a surrogate exist. The most common definition is an intermediate response measurement that accurately predicts the clinical endpoint of interest. This definition lacks some necessary details about what `accurately predicts' means and in what setting this prediction is done, but it is a useful starting point. \citet{Gilbert08b} refined the concept of a surrogate in the vaccine trial setting, proposing three levels: correlates, specific surrogates of protection, and general surrogates of protection. A correlate is a biomarker that is associated with the outcome in the vaccine arm or over both arms and a specific surrogate of protection, is a biomarker that is associated with efficacy. The definition of a general surrogate of protection is most closely related to the common definition of a surrogate: a biomarker on which the treatment effect can be used to accurately predict the treatment effect on the clinical outcome in a new setting. These levels are informative for general clinical trials as well. To make clear that we are considering the more general clinical trial setting, we will call biomarkers that are predictive of the clinical treatment effect in a new setting general surrogates (GS). Numerous papers have investigated the evaluation of biomarkers as GS, including \citet{Daniels97, Gail00, Burzykowski01, Dai12}.

The definition of a GS in \citet{Gilbert08b} does not define in what new trial settings the GS will be useful or if the surrogate is at the individual or trial level. An individual level surrogate is a measurement the treatment effect on which is predictive of the clinical effect at the individual level. The effect of treatment on a trial level surrogate can be used to infer what the trial results would have been had the clinical outcome been measured. A GS can be predictive at the trial level or the individual level, or both. As is outlined in \citet{Korn05}, these two types of surrogacy do not imply each other and can be unrelated. Most of the literature focuses on the evaluation of a potential trial level general surrogate in studies in which both the GS and the clinical outcome are measured, yet little attention is given to what information is available to support the generalizability of the association of the surrogate and the clinical endpoint to a new setting.  All existing general surrogate evaluation methods, to our knowledge, only consider summary measures for surrogate evaluation rather than comparison of candidate GS, and most are based on within sample prediction intervals. Only \citet{Baker06} attempts to formalize the exogenous quantification of the expected error when predicting in a new setting, using a similar summary of prediction error to our suggested method for a binary surrogate. In addition, most existing methods are specific for the type of data collected and require strict modeling assumptions. \citet{Dai12} propose a general method for evaluating a trial level GS, but they require individual level data be available in all trials and suggest a within sample evaluation of prediction error based on a linear relationship between the treatment effects. 

In this paper, we provide a precise definition of a trial level general surrogate. We propose a general and flexible evaluation and prediction method that differs from existing meta-analytic evaluation methods in several ways. Our suggested estimation method can be used on individual level or trial level data, similar to \citet{Daniels97} (DH). It allows for a flexible association between the treatment effect on the trial level general surrogate and the treatment effect on the clinical outcome and a flexible distribution for both true treatment effects over the trials. Our proposed evaluation method uses a Bayesian cross-validation approach to quantify the prediction error when estimating the treatment effect in a new setting. We propose the use of absolute prediction error, similar to that of \citet{Tian07} as a summary error for evaluation and comparison of candidate trial level general surrogate. The absolute prediction error is easily interpretable because it is on the scale of the treatment effect on the clinical outcome. We also outline a suggested nomenclature for classifying support for the generalizability of an evaluated trial level general surrogate. We build on the work of \citet{Daniels97}, taking a Bayesian approach to trial level general surrogate evaluation while allowing for the within trial estimation approach of \citet{Dai12} when individual level data are available. Software implementing our method in R and JAGS is available in the supplementary materials. 

We use our method to evaluate and compare trial level general surrogates for the pentavalent rotavirus vaccine RotaTeq\texttrademark (RV5) (Merck \& Co. Inc., Kenilworth, New Jersey).  RV5  has been shown to be efficacious against rotavirus gastroenteritis in many settings, but in developing countries where the disease burden is highest, this efficacy is lower than in developed nations. A universal and clear measure of post-randomization immune response that explains the efficacy differences among the trials and that could potentially accurately predict efficacy in future settings has yet to be clearly identified. We use our approach to attempt to identify trial level general surrogates in the setting of the RV5 trials.  

In Section \ref{Meth} we outline our proposed methods for estimation, prediction, evaluation and comparison. In Section \ref{Sim} we explore the operating characteristics of our methods in several settings, one of which follows closely the setting of the RV5 trials. In Section \ref{Gen} we outline our suggested nomenclature for the generalizability of a TLGS. In Section \ref{Rota} we use our method to investigate TLGS of rotavirus gastroenteritis of any severity in the RV5 trials, and in Section \ref{DIS} we summarize our findings and suggest possible extensions to the proposed methods. 

\section{Methods} \label{Meth}
We refine the definition of a GS from \citet{Gilbert08b} to indicate whether the prediction is at the individual or trial level and the characteristics of the new setting where such a GS might be used. If the trial level treatment effect on a biomarker can be used to predict (with low prediction error) a trial level clinical treatment effect and this predictive association is generalizable to a new setting, then the biomarker is a {\em trial level general surrogate} (TLGS) for that clinical outcome in that new setting.

This definition clearly states that the surrogate is at the trial level and indicates the settings where the TLGS can be used. 

To outline our proposed methods for evaluation and comparison of biomarkers as TLGS, we first define some notation. For subject $i \in \{1, \ldots,N_j\}$ in trial $j \in \{1,\ldots,J\}$, let $Z_{i,j}=\{0,1\}$ be the treatment indicator, $Y_{i,j}$  the clinical endpoint (the same over all trials) and ${A}_{i,j,k}$ the $k$th biomarker measure, $k \in \{1, \ldots, K\}$. At the trial level, let $T_{1,j,k}$ be the true treatment effect on the $k$th biomarker measurement and let $T_{2,j}$ be the true treatment effect on the clinical outcome. Let $N_{j,k}$ and $N_{j,2}$ be the set of subjects with the $k$th candidate TLGS and the clinical outcome measured in trial $j$, respectively. 
 
\subsection{Within Trial Models} \label{data} 
\subsubsection{Models for Subject Level Data} \label{WIT}
When subject level data are available in one or more trials, we  specify the following linear models for the (transformed) mean, median, or hazard for the  candidate TLGS and the clinical outcomes,
\begin{eqnarray}
 h_{1,k}(F[A_{i,j,k}|Z_{i,j}])&=&\zeta_{0j,k} + T_{1,j,k} Z_{i,j}   \label{T1mod},\\
 h_2(F[Y_{i,j}|Z_{i,j}])&=&\gamma_{0j} + T_{2,j} Z_{i,j}  \label{T2mod},
\end{eqnarray}
where $\zeta_{0j,k}$ and $\gamma_{0j}$ denote the $j$th trial specific intercepts for the $k$th biomarker measurement and the clinical outcome, respectively. Models (\ref{T1mod}) and (\ref{T2mod}) are specified in terms of the distribution function, $F(.)$, of A and Y and can be any set of models from which asymptotically normal coefficient estimates are obtained, including generalized linear models and  survival models. We fit the models separately for each trial and response (outcome or biomarker). For this reason, $N_{j,k}$ and $N_{j2}$, the subjects in each trial used to fit (\ref{T1mod}) and (\ref{T2mod}) respectively, do not need to have both the outcome and biomarkers provided the (implicit) missingness is random and/or by design.   
We denote the estimates of the treatment effects as  $\widehat{T}_{1,j,k}$ and $\widehat{T}_{2,j}$ with standard errors, $\hat{\sigma}_{1,j,k}$ and  $\hat{\sigma}_{2,j}$,  $j \in \{1, \ldots, J\}$.  

Let $\mathbf{O_{J,k}}$ be the vector of estimated treatment effects and estimated standard errors over all $J$ trials for the $k$th candidate GS; $\mathbf{O_{J,k}} = (\widehat{T}_{1,1,k} \ldots \widehat{T}_{1,j,k}, \widehat{T}_{2,1} \ldots \widehat{T}_{2,J}, \widehat{\sigma}_{1,1,k} \ldots \widehat{\sigma}_{1,j,k},  \widehat{\sigma}_{2,1} \ldots \widehat{\sigma}_{2,J})$ and define $\mathbf{O_{J}}$ to be the vector of all estimated treatment effects and standard errors over all $K$ candidate GS and clinical outcome data in a given set of $J$ trials. We will call $\mathbf{O_{J}}$ the vector of observed 'data'. Also define $\mathbf{T}_{2}=(T_{2,1} \ldots T_{2,J})$ and $\mathbf{T}_{1,k}=(T_{1,1,k} \ldots T_{1,J,k})$. 
 
\cite{Dai12} use an estimating equation approach to estimate the joint sampling distribution of the estimated treatment effects from (\ref{T1mod}) and (\ref{T2mod}). Using their approach the correlation between the estimated treatment effects can be  estimated. However, an estimate of this covariance is not required to evaluate the candidate TLGS (\citep{Daniels97}), and complex missingness by design can make this covariance estimation complex.

\subsubsection{Approximate Likelihood} \label{BTM}
We assume that the estimated treatment effects  are consistent and asymptotically normal and can be well approximated by 
\[
\widehat{T}_{1,j,k} \approx N(T_{1,j,k}, \hat{\sigma}^2_{1,j,k}) \hspace{3mm}\mbox{and}\hspace{3mm} \widehat{T}_{2,j} \approx N(T_{2,j}, \hat{\sigma}^2_{2,j}). \label{WCIL}
\]
Note that the estimated treatment effects can be obtained by fitting models (\ref{T1mod}) and (\ref{T2mod}) or from the literature.  
We denote these models for the estimated treatment effects given the true treatment effects by $f_{1jk}(\widehat{T}_{1,j,k}|T_{1,j,k},\hat{\sigma}_{1,j,k})$ and $f_{2,j}(\widehat{T}_{2,j}|T_{2,j}, \hat{\sigma}_{2,j})$. 
The joint distribution of the estimated treatment effects is given by
\begin{eqnarray}
&&f_{j,k}(\widehat{T}_{1,j,k},\widehat{T}_{2,j}|T_{1,j,k},T_{2,j},\hat{\sigma}_{1,j,k}, \hat{\sigma}_{2,j})=
f_{1jk}(\widehat{T}_{1,j,k}|T_{1,j,k},\hat{\sigma}_{1,j,k}) \times f_{2,j}(\widehat{T}_{2,j}|T_{2,j},\hat{\sigma}_{2,j}), \label{WCIL}
\end{eqnarray}
where we assume independence conditional on the true treatment effects. 
Based on these (approximate) trial specific models, the  likelihood over all the trials for the $k$th TLGS is given by $\prod^{J}_{j=1}{f_{j,k}(\widehat{T}_{1,j,k},\widehat{T}_{2,j}|\hat{\sigma}_{1,j,k}, \hat{\sigma}_{2,j},T_{2,j},T_{1,j,k})}$, as the estimates between trials are assumed to be  independent. In what follows, the estimated standard errors $\hat{\sigma}_{1,j,k}$ and $\hat{\sigma}_{2j,k}$ will be treated as fixed. 

When the individual level data is available, we can estimate the covariance of $(\widehat{T}_{1,j,k}, \widehat{T}_{2,j})$, and replace the two independent normals with a bivariate normal.
One could also use the individual data directly by using models  (\ref{T1mod}) and (\ref{T2mod}) to form the likelihood. However, such a formulation is unlikely to substantially improve evaluation of GS at the trial level as was demonstrated in \citet{Korn05}. Therefore, we focus on a two step approach: first, estimating the treatment effects within each of a set of trials, and second, constructing an approximate likelihood based on those estimates. 

\subsection{Between Trial Model and Priors} \label{prior} 
We specify a non-parametric prior for the distribution of the true treatment effect on the candidate GS, $T_{1,j,k}$. In particular we propose a Dirichlet process mixture of normals \citep{maceachern98},
\begin{eqnarray} 
T_{1,j,k} &\sim& N(\mu_{1,j,k},\tau_{1,j,k}) \\  \nonumber
({\mu_{1,j,k}}, {\tau_{1,j,k}}) &\sim& G\\ \label{T1prior}
G&\sim&DP(\omega, G_0) \nonumber
\end{eqnarray} 
For $\omega$,  we specify a $U(1,J)$ prior. We use this prior both to bound $\omega$ away from zero and to allow for clustering in the true trial level biomarker effects, $T_{1,j,k}$. We specify the base measure $G_0$  to be the product of a  normal and a gamma distribution where each $\mu_{1,j,k}$ follows a normal distribution with mean $\varphi_{1,k}$ and variance $\sigma^{2}_{\mu_{1,k}}$ and each $1/\tau_{1,j,k}$ follows a  gamma distribution with parameters $(\psi_{1,k}, \xi_{1,k})$. 
Denote the vectors of trial level parameters as $\boldsymbol{\mu_{1k}}=(\mu_{1,1,k}, \dots, \mu_{1,J,k})$ and $\boldsymbol{\tau_{1k}}=(\tau_{1,1,k}, \dots, \tau_{1,J,k})$. 
We use data-dependent hyper-priors for the parameters in the base measure as recommended in \citet{Taddy08}. 

We now specify a flexible model for the true trial level clinical treatment effect given the true trial level treatment effect on the TLGS, $[T_{2,j}|T_{1,j,k}]$. 
Specifically, $T_{2,j}=m(T_{1,j,k}, \beta_k, b_k) + \epsilon_{j,k}$, where $\epsilon_{j,k}$ are independent $N(0,\sigma^{2}_{\epsilon})$. The function $m(T_{1,j,k}, \beta_k, b_k)$ is defined as:
\begin{eqnarray}
m(T_{1,j,k}, \beta_k, b_k) = \beta_{0,k} + \beta_{1,k} T_{1,j,k}  + \sum^{M}_{m=1}{b_{m,k}|T_{1,j,k}-r_{m}|} \label{spline}
\end{eqnarray}
The $r_{1}<r_{2}<\ldots<r_{M}$ are fixed knots/changepoints. The coefficients associated with each knot, $b_{m,k}$ are penalized/shrunk using 
$N(0, \kappa^{2}_{b_{k}})$, 
with 
$ \kappa^{2}_{b_{k}}$ given an inverse gamma$(1,3)$ prior 
as recommended in \citet{Crainiceanu07}. 
The parameters $\beta_{0,k}$ and $\beta_{1,k}$ are given independent $N(0, \sigma^{2}_{\beta_k})$ priors,
with $1/\sigma^{2}_{\beta_k}$ set to 1$e^{-6}$. 
We use a diffuse inverse-gamma prior for $\sigma^{2}_{\epsilon}$.
The full set of hyper-parameters will be denoted as $\boldsymbol{\nu}=\{\boldsymbol{\tau^{2}}_{1k},\boldsymbol{\mu}_{1k}, \varphi_{1,k}, \sigma^{2}_{\mu_{1,k}},\psi_{1,k}, \linebreak[0] \xi_{1,k},\linebreak[0] \mathbf{b_k},\linebreak[0] \kappa^{2}_{b_{k}}, \linebreak[0] \sigma^{2}_{\epsilon},  \linebreak[0] \beta_{0,k},\linebreak[0] \beta_{1,k}, \linebreak[0]\omega\}$.

For the evaluation procedure presented in Section \ref{Eval},  we also need the null distribution of $\mathbf{T_2}$, i.e., the distribution of the $\mathbf{T_2}$ independent of $\mathbf{T_{1}}$. We do not use the marginal $\mathbf{T_2}$ derived from the above models to avoid any potential model misspecification for the model for $\mathbf{T_2} | \mathbf{T_1}$.  Instead, we specify the (null) prior for the true clinical treatment effects $\mathbf{T_2}$ ignoring all candidate TLGS information using the same specification and priors as in (\ref{T1prior}).

To compare candidate trial level surrogates, more than one candidate will need to be available in a given set of completed trials. When $K$ candidates are available, we use an additive model for $[T_{2,j}|T_{1,j,k}, \ldots T_{1,j,K}]$,  $T_{2,j}=\sum_k m_k(T_{1,j,k}, \beta_k, b_k) + \epsilon_j$, where $m$ has the form as (\ref{spline}). The model containing all information from all $K$ candidates is called the {\em full model}.

\subsection{Evaluation and Comparison} \label{Eval}
Our objective is to predict the true clinical treatment effect in a new setting $J+1$ where the estimated treatment effect on the clinical outcome, $\widehat{T}_{2,J+1}$, is not observed. For this prediction we use the posterior distribution of $T_{2,J+1}$, the true treatment effect on the clinical outcome in the new trial $J+1$, given the observed data, which includes  $\widehat{T}_{1J+1,k}$ and $\hat{\sigma}_{1J+1,k}$ in addition to $\mathbf{O_{J,k}}$, i.e., $T_{2,J+1}|\mathbf{O_{J,k}},\widehat{T}_{1,J+1,k}, \widehat{\sigma}_{1,J+1,k}$. As we are interested in comparing and evaluating candidate TLGS based on their predictive power, we need to determine the quality of the predictions from this distribution. 

To evaluate predictive accuracy and compare candidate TLGS, we will use the expected absolute prediction error,
\begin{eqnarray}
 D_{J+1,k} = E|T_{2,J+1} - \widehat{T}^{*}_{2,J+1}|, \label{D}
 \end{eqnarray} 
where $\widehat{T}^{*}_{2,J+1}$ is our point prediction of the treatment effect on the outcome in the new setting (trial $J+1$).  
The summary $D_{J+1,k}$ is the expected absolute prediction error for a new trial setting $J+1$. The merits and properties of the expected absolute prediction error were outlined in \citet{Tian07} for independent and identically distributed data where the true outcome, $T_{2}$, was observed. In our setting, the $T_{2}$ are unlikely to be identically distributed, although we assume that they are independent, and they are never observed. To estimate $D_{J+1,k}$, we could estimate the error we make for each observed trial $j$ by comparing a leave-one-out estimate, $\widehat{T}^{*}_{2,j}$ to the true $T_{2,j}$ value, $\widehat{d}_{j,k}$,  and then average over the $J$ trials, 
\begin{eqnarray}
 \widehat{D}_{J+1,k} = (1/J)\sum_j \widehat{d}_{j,k} = (1/J)\sum_j |T_{2,j} -\widehat{T}^{*}_{2,j}| \label{Dhat}.
 \end{eqnarray} 
A cross-validated estimate of this nature was found to have good finite sample size properties for iid data in \citet{Tian07}. Recall though, that we do not observe the true $T_{2,j}$, so we cannot directly use (\ref{Dhat}) either.

Our leave one out prediction of $T_{2,j}$, $\widehat{T}^{\star}_{2,j}$, is computed from the posterior obtained by excluding 
$\widehat{T}_{2,j}$ and $\widehat{\sigma}_{2,j}$ as well as all information about all other candidate GS, from the data, 
$$\widehat{T}^{\star}_{2,j} = E[T_{2,j}|\mathbf{O_{J(-j),k}}, \widehat{T}_{1,j,k}, \widehat{\sigma}_{1,j,k}],$$ 
where $\mathbf{O_{J(-j),k}}$ is all information about the clinical outcome and the $k$th candidate GS in all $\{1,\dots,J\}$ trials less trial $j$. 
Since we do not observe the true $T_{2,j}$, even in the evaluation trials, we will use an approximation to the distribution of the absolute prediction error to assess the TLGS. In particular, we use
\begin{eqnarray}
\tilde{d}_{j,k}|\mathbf{O_{J}} \sim |T_{2,j} - \hat{T}^{\star}_{2,j}|,
\label{Dtil}
\end{eqnarray} 
the posterior distribution of the approximate absolute prediction error for trial $j$ using all the data, $\mathbf{O_{J}}$. Here $\hat{T}^{\star}_{2,j}=E[T_{2,j}|\mathbf{O_{J(-j),k}},\widehat{T}_{1,j,k}, \widehat{\sigma}_{1,j,k}]|$. By taking the expected value of (\ref{Dtil}), i.e., with respect to the posterior distribution, $T_{2,j}|\mathbf{O_{J}}$, we obtain an approximation of $\widehat{d}_{j,k}$ in (\ref{Dhat}). For evaluation of a TLGS in an arbitrary new setting, we use the mixture distribution, $\tilde{D}_{J+1,k} \sim (1/J) \sum_j \tilde{d}_{j,k}$. The expectation of $\tilde{D}_{J+1,k}$ is an approximation of (\ref{Dhat}) and  is a single value comparison for candidate TLGS in a new setting.  

The distribution of $\tilde{D}_{J+1,k}$ is the distribution of interest for comparing candidate TLGS and as $\tilde{D}_{J+1,k}$ is on the scale of the true clinical effect, a first step in TLGS evaluation can be a comparison of $\tilde{D}_{J+1,k}$ to the smallest clinically relevant effect, as was suggested in \citet{Baker06}. However, to evaluate the absolute quality of a candidate TLGS we will compare  $\tilde{D}_{J+1,k} $ to the absolute prediction error in the absence of $\widehat{T}_{2,j}$ or $\widehat{T}_{1,j,k}$; we will denote this as $\widehat{D}_{J+1,0}$. 
We use the null model prior for the distribution of the true $T_{2,j}$ introduced at the end of Section \ref{prior}.  
Similar to the development for $\widehat{d}_{j,k}$, we use the following approximate distribution of the absolute prediction error,
\begin{eqnarray}
\tilde{d}_{j,0} |\mathbf{O_{J}} \sim |T_{2,j} - E[T_{2,j}|\mathbf{O_{J(-j, -1)}}] |, \label{D0}
 \end{eqnarray} 
where $\mathbf{O_{J(-j, -1)}}$ denotes the observed clinical outcome information less that in trial $j$, i.e., $\mathbf{O_{J(-j, -1)}} =(\widehat{T}_{2,q}, \widehat{\sigma}_{2,q}): q \in \{1, \ldots, j-1, j+1, \ldots, J\}\}$. 
The distribution $\tilde{D}_{J+1,0} \sim (1/J)\sum_j \tilde{d}_{j,0}$. $\tilde{D}_{J+1,0}$  is a good baseline for candidate TLGS evaluation because it quantifies the amount of error that occurs when no potential TLGS are used in the model. 
 
\begin{figure}
\centering
\includegraphics[scale=.75]{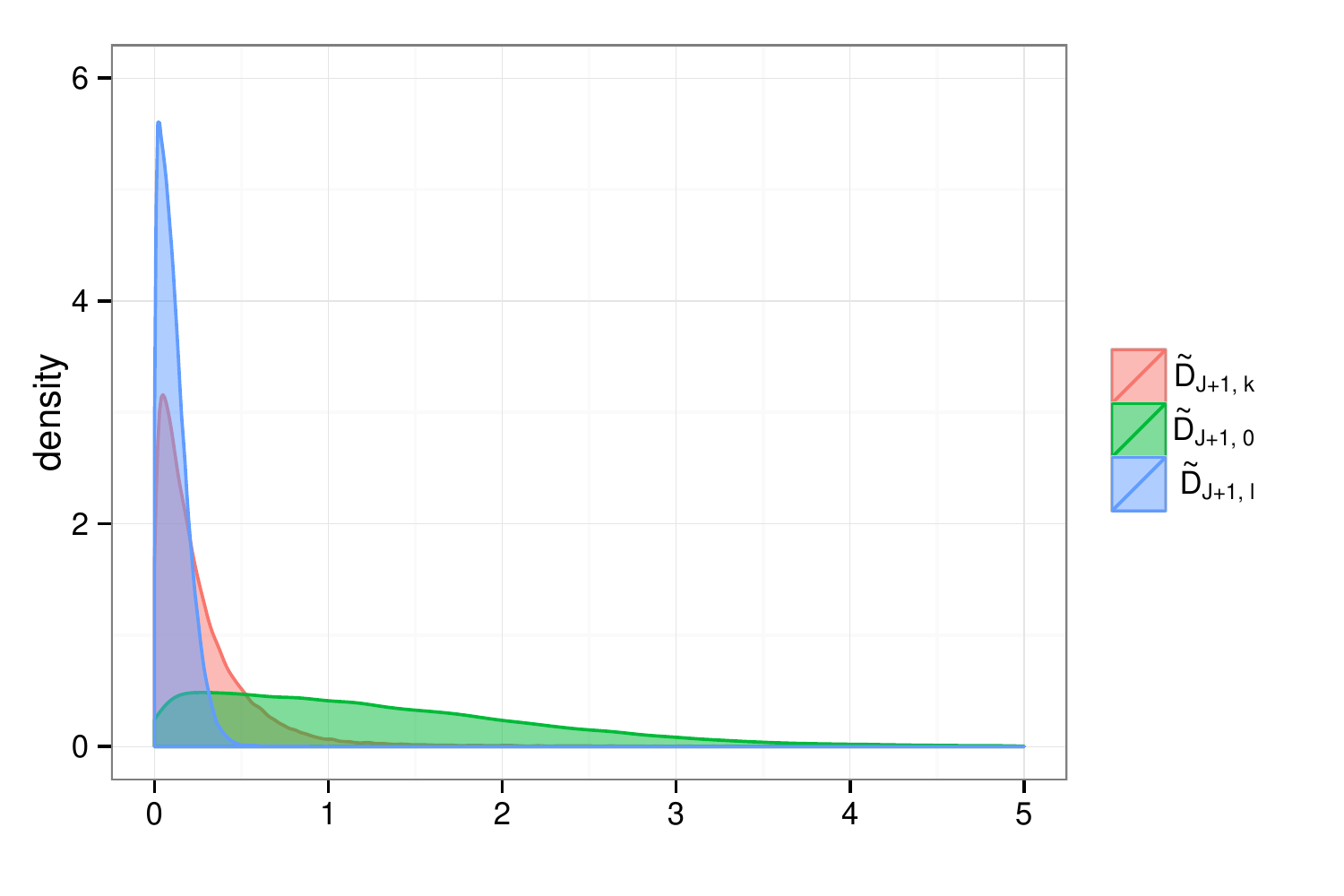}
\caption{Density estimates for $\tilde{D}_{J+1,k}$, $\tilde{D}_{J+1, l}$ and $\tilde{D}_{J+1, 0}$.  Candidate TLGS $k$ is a good, but not perfect GS. Candidate $l$ is an ideal TLGS, with a true $\widehat{D}_{J+1,l}=0$. 
$\tilde{D}_{J+1,0}$ corresponds to the model containing no candidate GS\label{density}.  Large mass closer to 0, and less spread away from zero, for $\tilde{D}_{J+1,k}$ and $\tilde{D}_{J+1,l}$, suggests a smaller  mean and lower variance of the approximate distribution of the absolute prediction error.}
\end{figure}

To evaluate and to compare candidate GS based on $\tilde{D}_{J+1,k}$ and $\tilde{D}_{J+1,0}$ , a joint graphical representation of the density estimates can be useful.  Densities of  $\tilde{D}_{J+1,k}$ that have more mass at 0 with shorter tails are evidence of superior candidates, as this implies greater probability of lower prediction error. Figure \ref{density} is an illustration of how the density plots can be used to compare and to evaluate candidate trial level general surrogates. Both candidate TLGS in Figure \ref{density} are superior to the null, $\tilde{D}_{J+1,0}$, while the density of $\tilde{D}_{J+1, l}$ in comparison to that of $\tilde{D}_{J+1,k}$ suggests that candidate $l$ is superior to candidate $k$; in this example, candidate $l$ is an ideal TLGS, with a true $\widehat{D}_{J+1,l}=0$, from using the true $T_{2,j}$ as the estimated treatment effect on the candidate GS. However, the $k$th candidate is still a very useful TLGS.  

The probability $P(\tilde{D}_{J+1,0}<\tilde{D}_{J+1,k})$, can be used to quantify the strength of evidence that a given candidate TLGS has any value. Small probabilities suggest that there is evidence that the $k$th candidate has value as a TLGS to aid in the prediction of the clinical treatment effect in a new trial setting. One can also rank candidates using the set of point and interval estimates from the set of $\{\tilde{D}_{J+1,1},\ldots,\tilde{D}_{J+1,K}\}$ distributions. The probability $P(\tilde{D}_{J+1,k}<\tilde{D}_{J+1,l})$, can be used to quantify the strength of evidence for the superiority of a given candidate over another within the same set of evaluation trials. Here, small probabilities suggest that there is evidence of the superiority of the $l$th TLGS candidate over the $k$th candidate, while large probabilities provide evidence of the opposite. 

\subsection{Posterior computation}
Computation of all quantities of interest requires four steps, which we outline below.  
 \begin{enumerate}
\item Sample from the posterior of the full model using all observed data to obtain the marginal posterior of each $T_{2,j}$ in JAGS.  This step provides the best estimate of the 'true' value of the treatment effect on the clinical outcome. 
\item Removing the observed clinical data information for trial $j$,
we sample from the corresponding posterior in JAGS to obtain the leave-the-$j$th-trial-out-estimate, $T^{*}_{2,j}$. We repeat for all $J$ evaluation trials.
\item Using only the clinical information from the other trials besides trial $j$ and removing all candidate surrogate information, we sample from the corresponding posterior to 
obtain an estimate of the true treatment effect on the clinical outcome,
the marginal or null leave-the-$j$th-trial-out-estimate. We repeat this for all $J$ evaluation trials.
\item We use the posteriors of $\mathbf{T_2}$ from steps 1 and 2 to obtain the distribution of  $\tilde{D}_{J+1,k}$ and the posterior distributions of $\mathbf{T_2}$ from steps 1 and 3 to obtain the distribution  $\tilde{D}_{J+1,0}$.  
 \end{enumerate}

We use the posterior from step 2 to estimate the true clinical treatment effect in a trial setting where only the candidate surrogate is measured. 

\section{Simulations} \label{Sim} 
We investigate the performance of our method in several scenarios, including for different relationships between the true treatment effect on the clinical outcome, $T_{2,j}$, and the true treatment effect on the surrogate, $T_{1,j,k}$ and for different distributions governing the true $T_{1,j,k}$, including a single normal or a mixture of normals. For each of these scenarios we investigate the impact of the magnitude of the  errors with which the treatment effects were estimated within the trials, $\hat{\sigma}_{1,j,k}$ and  $\hat{\sigma}_{2,j}$. For the scenarios where $T_{1,j,k}$ are generated from a normal distribution, we fix the mean and standard deviation (SD) of each $T_{1,j}$ to be 2 and 1, respectively. In the mixture of normals settings, we generate $T_{1,j,k}$ from an equally weighted mixture of 3 normals with means $(-1,0,2)$ and SDs of 1. We fix the SD of the generated $T_{2,j}$, conditional on $T_{j,1,k}$, to be 1. 

For each scenario we simulate 200 datasets of 20 trials, unless otherwise specified. The exact specifications of the simulation scenarios are given in Appendix A of the supplementary materials; the R and JAGS code used to generate the data and fit the models is also available in the supplementary materials. The last scenario in each  table is based on the rotavirus vaccine trial application, with 12 trials and using the set of $\widehat{T}_1$ and $\widehat{\sigma}_1$ from one of the candidate GS as well as assuming a strong linear association between the true treatment effects. 
For our method outlined in Section 2, we use a piecewise linear model as given in (\ref{spline}) with knots at the 33rd and 66th percentiles of the $\widehat{T}_2$, for the prior on ${T_{2,j}}|{T_{1,j,k}}$.
We compare our method's predictive accuracy, as described in Section 2.4, with: 1) a linear regression on the {\em estimated} treatment effects (referred to as 'linear model' in the tables) and 2) the method of \citet{Daniels97} (referred to as DH in the tables). 

Table \ref{simset2} reports the mean and SD of $\widehat{D}_{J+1,k} = (1/J) \sum_j |T^{true}_{2,j}- E[T_{2,j}|\mathbf{O_{J(-j)}},\widehat{T}_{1,j,k}, \widehat{\sigma}_{1,j,k}]|$. $T^{true}_{2,j}$ is the true clinical treatment effect  for trial $j$ and $J$ is the number of trials. Table \ref{simset2} also displays the mean and bias of the posterior mean of $\tilde{D}_{J+1,k}$, given in (\ref{Dtil}), in comparison to  $\widehat{D}_{J+1,k}$. Both the DH method and our proposed method estimate $\widehat{D}_{J+1,k}$ more accurately than the linear model in all scenarios. The DH method and our proposed method have similar bias, both well below Monte Carlo error; this suggests relative unbiasedness for both methods. The $\widehat{D}_{J+1,k}$ are always lower for our proposed method in non-linear scenarios and similar in linear scenarios. 

Table \ref{simset3} reports the average ${P(\tilde{D}_{J+1,0}<\tilde{D}_{J+1,k})}$ over the 200 simulated datasets for each scenario and method. The averages are near $1/2$ for both the DH method and our proposed method in the first row of Table \ref{simset3}, the scenario where there is no useful candidate surrogate.  In addition, the lower probabilities in all other scenarios suggest that this probability can help quantify the value of a candidate as a TLGS. The low probabilities for all scenarios for the linear model are caused by the biased estimation of $D_{J+1,k}$, which makes $P(\tilde{D}_{J+1,0}<\tilde{D}_{J+1,k})$ under this method non-informative for evaluating TLGS candidates. Table \ref{simset3} also presents the average probability for comparing two candidates, $P(\tilde{D}_{J+1,k}<\tilde{D}_{J+1,l})$, where 
$\tilde{D}_{J+1,l}$ is the estimate based on a TLGS with a true $\widehat{D}_{J+1,l}=0$, from using $T^{true}_{2,j}$ as the estimated treatment effect on the candidate GS. The estimates of $P(\tilde{D}_{J+1,l}<\tilde{D}_{J+1,k})$ being larger than $1/2$  in almost every case suggests that this probability can be used to discern between candidate general surrogates of differing quality.

\begin{sidewaystable}
\begin{centering}
\caption{Mean (SD) $\widehat{D}_{J+1,k}$, Mean and Bias $\tilde{D}_{J+1,k}$: \label{simset2} $\widehat{D}_{J+1,k} = (1/J) \sum_j |T^{true}_{2,j}- E[T_{2,j}|O_{J(-j)},\widehat{T}_{1,j}, \widehat{\sigma}_{i,j}]|$ where $T^{true}_{2,j}$ is the true clinical treatment effect  generated for trial $j$.}
\begin{tabular}{l|ccc|ccc}
&\multicolumn{3}{c|}{Mean $\widehat{D}_{J+1,k}$ (SD)}& \multicolumn{3}{|c}{Mean $\tilde{D}_{J+1,k}$, Bias}\\
Scenario& Linear Model & DH& New Method& Linear Model & DH& New Method\\
0, Linear, no-effect & 0.87 (0.15) & 1.15 (0.15) & 1.28 (0.18) & 0.09, 0.65 & 1.15, 0 & 1.28, -0.002 \\
1, Linear, & 0.10 (0.018) & 0.47 (0.07) & 0.16 (0.025) & 0.012, 0.077 & 0.48, -0.005 & 0.16, 0.003 \\
2, Linear, large SE $\hat{T}_1$& 0.67 (0.11) & 1.06 (0.14) & 1.07 (0.17) & 0.07, 0.51 & 1.06, 0 & 1.07, 0.001 \\
3, Linear, large SE $\hat{T}_2$& 0.30 (0.11) & 0.73 (0.095) & 0.53 (0.098) & 0.089, 0.19 & 0.91, -0.13 & 0.59, -0.032 \\
4, Linear, large Beta& 0.09 (0.01) & 0.47 (0.08) & 0.14 (0.01) & 0.01, 0.064 & 0.47, -0.005 & 0.14, 0 \\
5, Linear, 10 trials& 0.13 (0.03) & 0.53 (0.12) & 0.23 (0.06) & 0.03, 0.09 & 0.54, -0.004 & 0.23, 0.003  \\
6, Linear, 5 trials& 0.19 (0.09) & 1.76 (3.79) & 0.58 (0.21) & 0.1, 0.076 & 1.76, -0.002 & 0.58, 0.005 \\
\hline
7,  Cubic-Threshold& 0.8 (0.26) & 1.08 (0.3) & 0.39 (0.21) & 0.12, 0.58 & 1.08, 0 & 0.39, 0.002  \\
8,  Step& 0.98 (0.12) & 1.16 (0.087) & 0.81 (0.16) & 0.11, 0.79 & 1.16, 0.001 & 0.81, 0.001 \\
9, Cubic&  0.4 (0.15) & 0.68 (0.18) & 0.18 (0.07) & 0.06, 0.30 & 0.68, -0.002 & 0.17, -0.001 \\
10, Linear, non-iid $\mathbf{T_1}$& 0.15 (0.028) & 0.39 (0.055) & 0.22 (0.039) & 0.016, 0.11 & 0.4, -0.004 & 0.22, 0.002  \\
11, Step, non-iid $\mathbf{T_1}$& 1 (0.13) & 1.13 (0.088) & 0.82 (0.22) & 0.11, 0.85 & 1.13, 0.003 & 0.82, 0.0  \\
12, Step, non-iid $\mathbf{T_1}$, 5 trials& 1.38 (0.6) & 2.58 (3.53) & 1.8 (0.68) & 0.69, 0.65 & 2.57, 0.008 & 1.79, 0.007  \\
\hline
13 Example scenario, 12 trials& 0.5 (0.09) & 0.79 (0.14) & 0.73 (0.16) & 0.093, 0.37 & 0.81, -0.009 & 0.68, 0.045 \\
\end{tabular}
\end{centering} 
\end{sidewaystable}

\begin{sidewaystable}
\begin{centering}
\caption{Detection of predictive power based on $\tilde{D}_{J+1,k}$  \label{simset3}: $\tilde{D}_{J+1,0}$ is the distribution of the prediction error for the null model containing no GS candidates.  $\tilde{D}_{J+1,k}$ is distribution of the prediction error  for the candidate created in each scenario, and $\tilde{D}_{J+1,l}$ is distribution of the prediction error for the  TLGS with a true $\widehat{D}_{J+1,l}=0$, using $T^{true}_{2,j}$ as the candidate TLGS. }
\begin{tabular}{l|ccc|ccc}
&\multicolumn{3}{c|}{$P(\tilde{D}_{J+1,0}<\tilde{D}_{J+1,k})$}& \multicolumn{3}{|c}{$P(\tilde{D}_{J+1,l}<\tilde{D}_{J+1,k})$}\\
Scenario& Linear Model & DH & New Method& Linear Model & DH & New Method\\
0, Linear no-effect & 0.033 & 0.424 & 0.452 & 0.963 & 0.675 & 0.944 \\
1, Linear & 0.005 & 0.227 & 0.075 & 0.636 & 0.424 & 0.603 \\
2, Linear large SE $\hat{T}_1$& 0.029 & 0.429 & 0.414 & 0.947 & 0.656 & 0.928 \\
3, Linear large SE $\hat{T}_2$& 0.038 & 0.374 & 0.255 & 0.197 & 0.467 & 0.536 \\
4, Linear large Beta& 0.004 & 0.222 & 0.064 & 0.624 & 0.418 & 0.577 \\
5, Linear 10 trials& 0.011 & 0.217 & 0.09 & 0.674 & 0.443 & 0.655  \\
6, Linear 5 trials& 0.022 & 0.25 & 0.147 & 0.527 & 0.496 & 0.669 \\
\hline
7,   Cubic-Threshold& 0.012 & 0.166 & 0.064 & 0.961 & 0.634 & 0.778  \\
8,  Step& 0.087 & 0.495 & 0.399 & 0.994 & 0.689 & 0.938 \\
9, Cubic&  0.036 & 0.45 & 0.127 & 0.895 & 0.542 & 0.559 \\
10, Linear, non-iid $\mathbf{T_1}$& 0.006 & 0.177 & 0.097 & 0.71 & 0.391 & 0.694  \\
11, Step, non-iid $\mathbf{T_1}$& 0.123 & 0.507 & 0.417 & 0.997 & 0.713 & 0.926  \\
12, Step, non-iid $\mathbf{T_1}$, 5 trials& 0.199 & 0.427 & 0.371 & 0.906 & 0.639 & 0.839  \\
\hline
13, Example scenario, 13 trials& 0.037 & 0.317 & 0.27 & 0.576 & 0.5 & 0.706 \\
\hline
\end{tabular}
\end{centering} 
\end{sidewaystable}

Tables S.2 and S.3 in the supplementary materials summarize the simulations using subject level data (as opposed to trial level summaries directly) to compare the two different ways of formulating the within-trial likelihoods. We find that the methods are basically comparable, with the working independence version of the likelihood having lower bias in the estimation of $\widehat{D}_{J+1,k}$ by $\tilde{D}_{J+1,k}$, although the bivariate normal likelihood method tends to have slightly smaller $\tilde{D}_{J+1,k}$ on average. 

\section{Generalizability} \label{Gen} 
We have outlined a method for evaluating and comparing candidate TLGS. Now we attempt to quantify the generalizability of a TLGS to a new setting. Suppose we have a new setting ($J+1$), where only the treatment effect on the TLGS is estimated. We can use our posited method to evaluate and compare candidate TLGS in the $J$ completed trials, then estimate the clinical treatment effect in this new setting. However, why should we assume that this estimate is valid in this new setting? If this new setting is similar enough to the $J$ settings where the TLGS was evaluated, we would have more confidence that the relationship between the TLGS and the clinical outcome will persist in this new setting.

Information to support generalizability can be  based on the characteristics of the new setting, such as the ethnic origin and age range of the trial subjects, relative to the evaluation settings. If information on the $J$ evaluation settings and the  new setting ($J+1$) is available to assess whether the new setting's and evaluation settings' characteristics are similar, it will strengthen generalizability. The characteristics that vary over the evaluation trial settings are often referred to as the units of variation. Although this is often pointed out in meta-analytic papers, the implications of this variation are typically not explored or discussed. 

Here, we suggest guidelines for conveying the information available to support the assumption of generalizability. In particular, we propose three ordered classes of generalizability support: \textit{represented}, \textit{within range} and \textit{outside the range}.  If the new trial setting is exactly the same as the evaluation setting in terms of available characteristics, findings are strongly generalizable. When all the characteristics of a new trial are present in at least one of the observed trials, but not all in the same one, we call the support of the generalizability assumption \textit{represented}. When not all characteristics of a new setting are represented in an observed trial, but all the characteristics are within the range of the observed trials, we call the support for the generalizability assumption \textit{within range}. An example of this third type of support is a new trial with participants between the ages 50 to 60 years, when previous trials enrolled 40-50 year old and 60-75 years old subjects. On the other hand, if a new setting only enrolled 20-30 year old  participants,
we would call the support for the generalizability assumption \textit{outside range}. Clearly the evidence to support generalization would be expected to decline from \textit{represented} to \textit{within range} to \textit{outside range}.  

The reliability of the above classifications hinges on observing a large number of (the same) characteristics in the evaluation settings and the new study. If there are very few observed characteristics in any of the $J$ evaluation settings or the new setting, the information to support generalizability is limited. The proposed nomenclature is solely a suggestion for succinctly conveying the information available to support the assumption of generalizability.

\section{Application: Pentavalent Rotavirus Vaccine} \label{Rota}
More than 450,000 children under five years died from complications of rotavirus infection each year \citep{Tate08} prior to vaccine availability. The pentavalent rotavirus vaccine RotaTeq\texttrademark (RV5) developed by Merck has been licensed for use in over 120 countries. The rotavirus Efficacy and Safety Trial (REST) against severe rotavirus gastroenteritis \citep{rota006} was conducted in 11 countries. In 
the substudy of REST in Finland and the United States  estimated efficacy was as high as 98\%. However, in other regions in Africa and Asia, lower efficacy has been observed \citep{rotaAfrica,rotaasia}. Lower efficacy may be related to difference in participants' immune system function.  And if this is the case, such measurements could be used to better predict efficacy in future settings. 

We investigate several candidate trial level general surrogates using data from four phase II and III studies of RV5 conducted in seven countries: Finland, the United States, Vietnam, Mali, Bangladesh, Ghana and Kenya \citep{rota006, rota005, rotaAfrica, rotaAfricaimmuno, rotaasia, rotaasiaimmuno}.   Participants from Finland and the United States were included in more than one study.  Some of the trials involved more than one  country. For the purpose of this analysis, we assume participants from different countries in the same trial to be independent trials as it is unlikely that outcomes will be correlated between countries even within the same trial setting. Table \ref{tab1} shows the 12 trials (based on study and country) used as our evaluation trials, and a 13th trial in Taiwan where only the potential TLGSs are measured. The data were provided by Merck \& Co. Inc., Kenilworth, New Jersey, through data sharing agreements with the Fred Hutchinson Cancer Research Center and the National Institute of Allergy and Infectious Diseases. 

For this application we consider rotavirus gastroenteritis of any severity as the outcome of interest. Dose, endemic burden of disease, age range, ethnicity, and region were available for each trial. Several candidate TLGS were measured in a subset of individuals.  These included serum anti-rotavirus IgA B-cell responses (IgA), as well as serum neutralization antibody (SNA) to the human rotavirus serotypes G1, G2, G3, G4 and P1A. 

Given that the different trials collected all the markers on only a random sample of those with the clinical outcome, we did not use the method of \citet{Dai12}, but instead present results using the working-independence model (\ref{WCIL}) based on estimates obtained from fitting independent generalized linear models in each trial for the available immunogenicity and clinical outcome data.

\begin{table}
\begin{centering}
\caption{Data available for each trial used in the analysis in Section 5.\label{tab1}}
\begin{tabular}{cccccc}
  \hline 
 &Region/ & &\multicolumn{3}{c}{Number of Subjects with}\\ 
Protocol Number & Grouping &Country & Clinical Outcome & IgA & G1 \\ 
  \hline
015 & Asia& Bangladesh & 1116 & 146 & 146  \\ 
 005$^{\dagger}$ & Finland  & Finland & 1027 & 644 & 647 \\ 
006 & Finland  & Finland & 2324 & 358 & 1503  \\ 
007 & Finland & Finland & 637 & 54 & 54  \\ 
015 & Africa& Ghana& 1971 & 143 & 143  \\ 
015 & Africa & Kenya& 1137 & 128 & 128  \\ 
015 & Africa & Mali& 1667 & 137 & 137 \\ 
 006 &Native lands& United States & 583 & 207 & 207 \\ 
 006 & US concomitant & United States & 1239 & 106 & 104  \\ 
 006 & US non-concomitant & United States & 366 & 220 & 210  \\ 
 007 & United States & United States & 478 & 97 & 98  \\ 
015 & Asia & Vietnam& 871 & 149 & 149 \\
006 & Asia& Taiwan & 0 & 99 & 99  \\ 
   \hline
\multicolumn{6}{l}{$\dagger$ Only the placebo and low, middle, and high dose RV5 groups were included.}
 \end{tabular}
\end{centering} 
\end{table}

\begin{figure}
\centering
\includegraphics[scale=0.75]{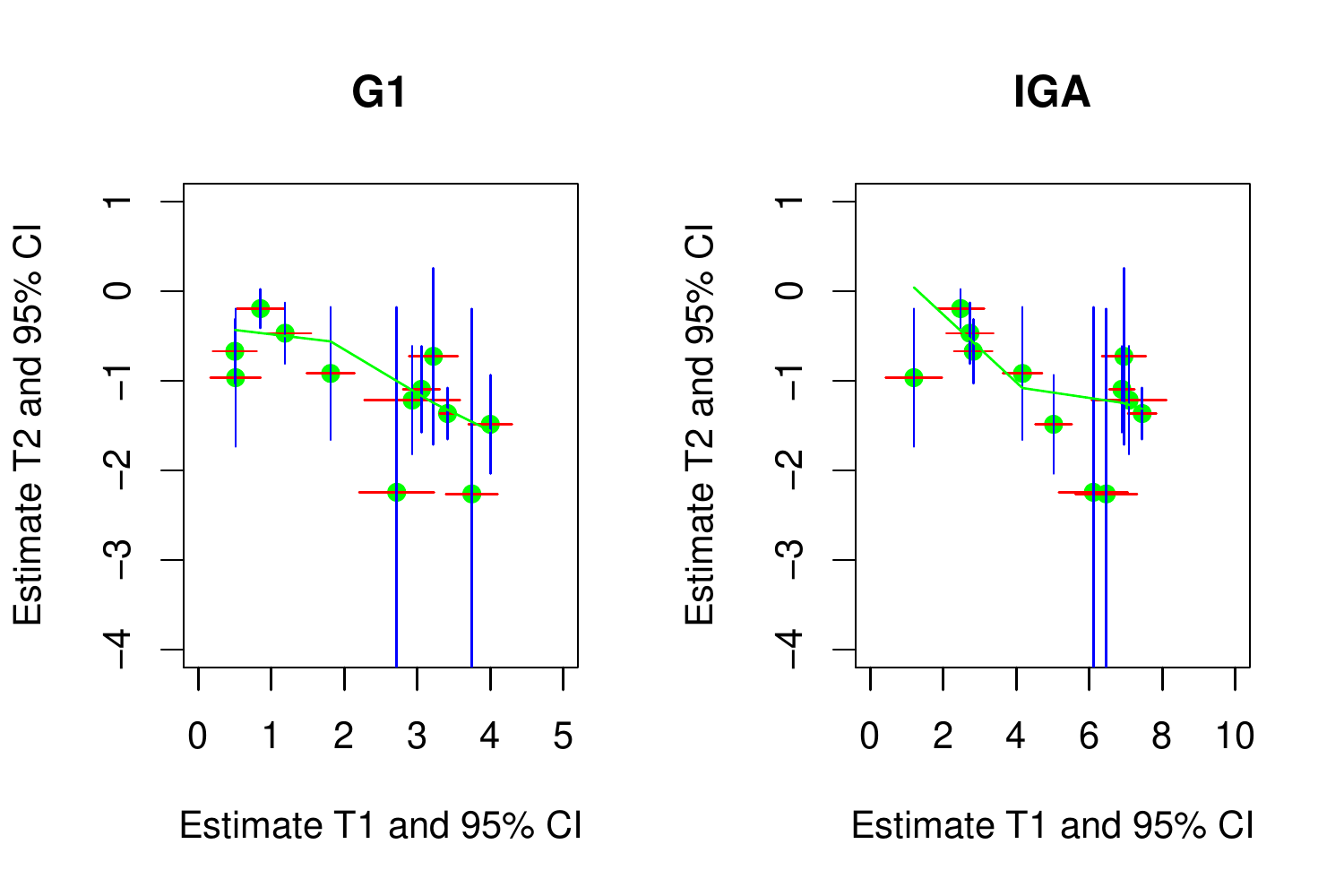}
\caption{Association of Estimated Vaccine Effects. \label{examplefig1} Each sub-figure displays the relationship between the estimated treatment effect on the candidate trial level GS, $\widehat{T}_{1,k,j}$, on the x-axis, and the estimated treatment effect on the clinical outcome, $\widehat{T}_{2,j}$, on the y-axis, over the 12 trial settings. The two candidates of interest are G1 or IgA. The crossbars depict the 95\% CI for the estimates. The smooth curve overlaid is the posterior mean from the spline model given in (\ref{spline}).}
\end{figure}

Figure \ref{examplefig1} displays the point estimates and 95\% credible intervals (CI) for the vaccine effect on the clinical outcome (rate of gastroenteritis of any severity), $T_{2,j}$ and the vaccine effect on two selected immune markers, SNA G1 and serum anti-rotavirus IgA, $T_{1,j,k}$. Due to the limited number of trial settings, we limited the number of candidates for comparison to the two candidates that seemed to have the best association with outcome as seen in Figure \ref{examplefig1}. Figure S1 in the supplementary materials describes this same relationship for all the other biomarkers collected. The smooth curve overlaid in all figures is the posterior mean from the regression model given in (\ref{spline}).

The $T_{2,j}$ estimates and standard errors are obtained by fitting Poisson regression models, including an offset for the log of follow-up time. This is the same model used in the clinical papers on these trials \citep{rota005, rota006, rotaAfrica, rotaAfricaimmuno, rotaasia, rotaasiaimmuno}. The clinical outcome in each trial is the difference in log rates of rotavirus gastroenteritis of any severity between the vaccinated and unvaccinated. A linear model was used to estimate the vaccine effect on the two potential TLGS with the effect being the difference in the log titer level between the vaccinated and unvaccinated participants.

\begin{figure}
\centering
\includegraphics[scale=0.75]{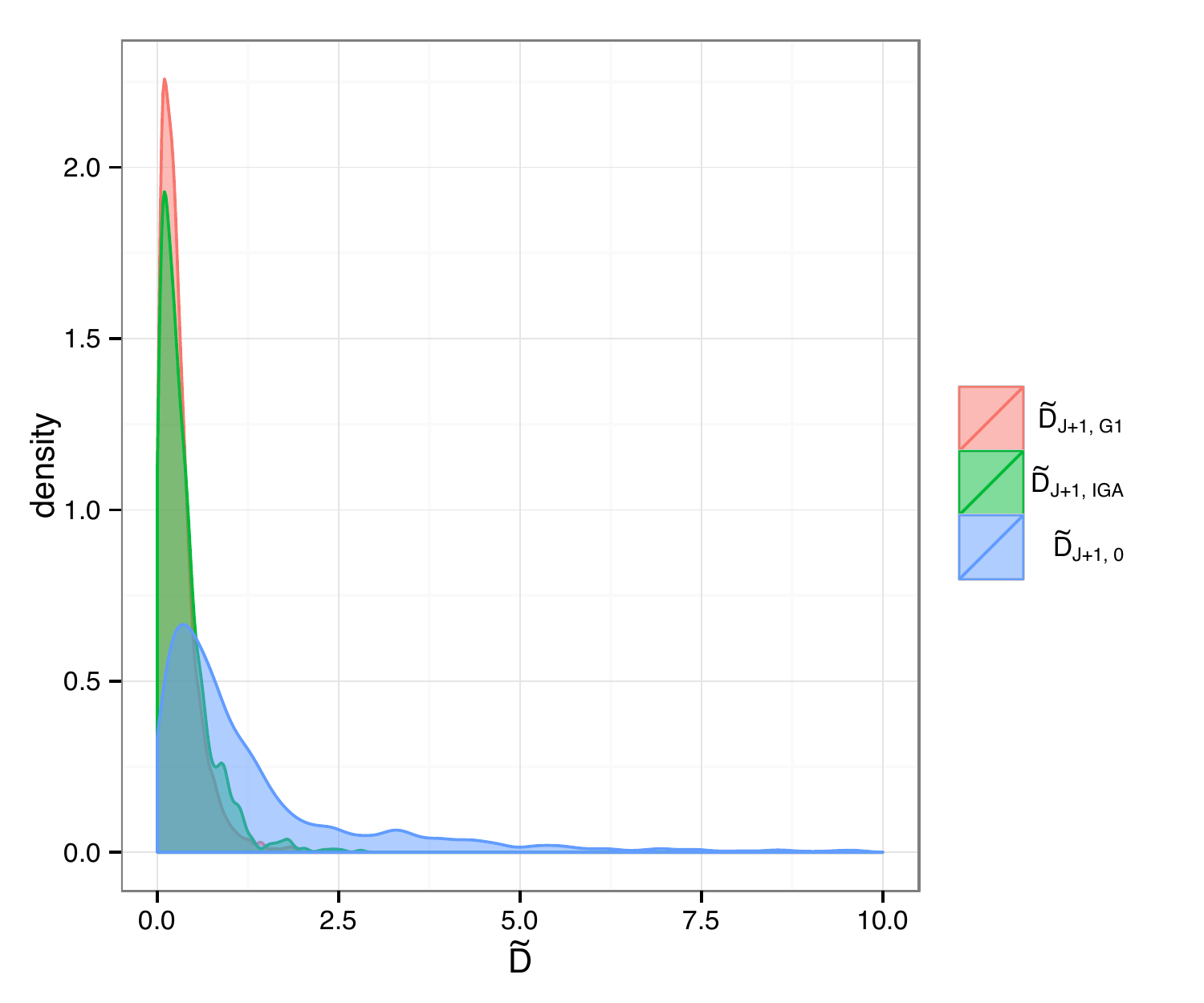}
\caption{Density estimates for $\tilde{D}_{J+1,IgA}$, $\tilde{D}_{J+1,G1}$  and $\tilde{D}_{J+1,0}$ \label{ExampleD}}
\end{figure}

For ease of notation let $k=\{IgA, G1\}$. 
The mean of the predictive error distribution, $D_{J+1,G1}$ is 0.29 with a 95th percentile of 0.8; for $D_{J+1,IgA}$, 0.36, with a 95th percentile of 1.03. We find evidence that both serum anti-rotavirus IgA and SNA G1 have value as trial level general surrogates for rotavirus gastroenteritis of any severity in settings where they are generalizable. This can be seen in Figure \ref{ExampleD} and from the probabilities $P(\tilde{D}_{J+1,0}<\tilde{D}_{J+1,G1})=0.19$ and  $P(\tilde{D}_{J+1,0}<\tilde{D}_{J+1,IgA})=0.22$ being less than $0.5$. The probability $P(\tilde{D}_{J+1,GI}<\tilde{D}_{J+1,IgA})=0.46$ suggests the two candidates are similar in quality. These results can also be seen in Figure \ref{ExampleD}, as both $\tilde{D}_{J+1,IgA}$ and $\tilde{D}_{J+1,G1}$ have more mass closer to 0 and less spread than $\tilde{D}_{J+1,0}$; there is also weak evidence that G1 is a slightly better TLGS than IgA as indicated by the probability $P(\tilde{D}_{J+1,GI}<\tilde{D}_{J+1,IgA})=0.46$ and by the higher peak of the $\tilde{D}_{J+1,G1}$ density that is close to 0.

Only selected sites had the clinical outcome measure collected in the REST clinical trial \citep{rota006}. Immune measurements alone were taken at the Taiwan site. We estimate the true clinical vaccine effect at the Taiwan site, $T_{2, Taiwan}$ (`Taiwan' corresponds to $J+1$) to be -1.51 based on the vaccine effect on G1 with 95\% CI $(-2.3,-0.90)$; this corresponds to a vaccine efficacy estimate of 78\% with 95\% CI (59\%, 90\%) against rotavirus gastroenteritis of any severity. Given the 95th percentile of the distribution $\tilde{D}_{J+1,G1}$ and the the upper 95\% CI limit of the credible interval for $T_{2, Taiwan}$ it is unlikely there would not have been positive efficacy in Taiwan as $(-0.9+0.8)<0$. Figure S2 of the supplementary materials illustrates this graphically.

The generalizability of SNA G1 as a TLGS to Taiwan can be classified between \textit{within range} and \textit{outside of range}; this trial was at the same dose, age range, burden of disease in the population, and general region, Asia, as previous trials. However, previous trials conducted in Asia were in Vietnam and Bangladesh, which are socioeconomically different from Taiwan. 

\section{Discussion} \label{DIS}
We have provided a definition of a TLGS and outlined a flexible Bayesian framework for the prediction of clinical treatment effects in a new setting, given the estimated treatment effect on a candidate TLGS. We also proposed a useful summary for the evaluation and comparison of candidate TLGS. We demonstrate that our prediction method generally has better predictive properties than previous methods, particularly when the true relationship between the treatment effects is non-linear. We also describe a nomenclature for conveying the evidence to support generalizability of a trial level general surrogate. 

In the application, we find evidence of two useful trial level general surrogates for rotavirus gastroenteritis of any severity in the RV5 vaccine trials; similar findings suggesting serum anti-rotavirus IgA as a surrogate have been presented by \citet{Goveia14}. We used the treatment effect on SNA G1 to predict the clinical efficacy of RV5 against rotavirus gastroenteritis of any severity in the Taiwan region and found that it is likely that there would have been positive efficacy observed in this region if the clinical endpoint had been collected. As is demonstrated in the application and pointed out in Section \ref{WIT}, our proposed method allowed us to use all available outcome and immunogenicity data in each trial to estimate the treatment effect on the clinical outcome and the treatment effect on the candidate GS. Unlike other methods, subjects need not have both the  candidate trial level general surrogate and clinical outcome data in each trial, provided the implicit missingness is not informative, as was the case in the RV5 trials. 

The evaluation and comparison methods we have developed can be used with any flexible modeling method as is demonstrated by the DH simulation results. Our proposed Bayesian non-parametric model could also be used for the full-data model estimation, and other models could be considered for the leave-the-$j$th-trial-out predictions, such as the DH model. Useful extensions to our proposed method would be the simultaneous evaluation of the individual level GS and TLGS, such as is discussed in \citet{Alonso14} or the evaluation of surrogate consistency, as discussed and demonstrated for a specific meta-analytic setting in \citet{Elliott14}. The consideration of combinations of measures as surrogates is also of interest for future research. 

\section*{Acknowledgments}
We are grateful to Merck \& Co.\ Inc.\ (Kenilworth, New Jersey) for sharing their RotaTeq\texttrademark ~~clinical trial data with us. The authors received no funding from Merck \& Co.\ Inc.\ for this analysis nor did Merck \& Co.\ Inc.\ influence the analysis in any way. 
M.J.\ Daniels was partially supported by National Institutes of Health R01 CA183854. M.E.\ Halloran was partially supported by National Institutes of Health 
R37 AI032042. 

\section*{Supplementary Materials}
The supplementary materials contain further data summaries and additional simulation results. 

\newpage

\bibliographystyle{plainnat}
\bibliography{bib_TLGS}

\end{document}